\documentclass[conference]{IEEEtran}
\IEEEoverridecommandlockouts
\usepackage{cite}
\usepackage{amsmath,amssymb,amsfonts}
\usepackage{algorithmic}
\usepackage{graphicx}
\usepackage{textcomp}
\usepackage{xcolor}
\usepackage{float}
\usepackage{dblfloatfix} 

\newcommand{\sate}{state-of-the-art}
\newcommand{\etal}{\textit{et al.}}

\def\BibTeX{{\rm B\kern-.05em{\sc i\kern-.025em b}\kern-.08em
    T\kern-.1667em\lower.7ex\hbox{E}\kern-.125emX}}
\begin{document}

\title{Preservation of High Frequency Content for Deep Learning-Based Medical Image Classification
}

\author{\IEEEauthorblockN{Declan McIntosh}
\IEEEauthorblockA{\textit{Electrical and Computer Engineering} \\
\textit{University of Victoria}\\
Victoria, Canada \\
declanmcintosh@uvic.ca}
\and
\IEEEauthorblockN{Tunai Porto Marques}
\IEEEauthorblockA{\textit{Electrical and Computer Engineering} \\
\textit{University of Victoria}\\
Victoria, Canada \\
tunaip@uvic.ca}
\and
\IEEEauthorblockN{Alexandra Branzan Albu}
\IEEEauthorblockA{\textit{Electrical and Computer Engineering} \\
\textit{University of Victoria}\\
Victoria, Canada \\
aalbu@uvic.ca}
}

\maketitle

\begin{abstract} 
    Chest radiographs are used for the diagnosis of multiple critical illnesses  (\textit{e.g.}, Pneumonia, heart failure, lung cancer), for this reason, systems for the automatic or semi-automatic analysis of these data are of particular interest. An efficient analysis of large amounts of chest radiographs can aid physicians and radiologists, ultimately allowing for better medical care of lung-, heart- and chest-related conditions. We propose a novel Discrete Wavelet Transform (DWT)-based method for the efficient identification and encoding of visual information that is typically lost in the down-sampling of high-resolution radiographs, a common step in computer-aided diagnostic pipelines. Our proposed approach requires only slight modifications to the input of existing state-of-the-art Convolutional Neural Networks (CNNs), making it easily applicable to existing image classification frameworks. We show that the extra high-frequency components offered by our method increased the classification performance of several CNNs in benchmarks employing the NIH Chest-8 and ImageNet-2017 datasets. Based on our results we hypothesize that providing frequency-specific coefficients allows the CNNs to specialize in the identification of structures that are particular to a frequency band, ultimately increasing classification performance, without an increase in computational load. The implementation of our work is available at github.com/DeclanMcIntosh/LeGallCuda.
\end{abstract}


\section{Introduction}

Chest radiograph testing for medical diagnosis is an important and widely used component of care for a multitude of afflictions. This is particularly important during the COVID-19 pandemic, as it has been shown by the Canadian Association of Radiologists \cite{cdc} that chest radiographs constitute valuable supplemental diagnosis methods for infections related to the virus. Chest radiographs also are commonly used for the diagnosis and treatment of conditions associated with, but not exclusively related to, the SARS-CoV-2 virus such as Pneumonia \cite{Speets933}. The need for efficient, timely and cost-effective automated detection of abnormalities in radiograph images expands beyond any singular outbreak as Pneumonia was the worldwide top cause of death for infants in 2017 \cite{owidpneumonia}. More generally, cardiothoracic and pulmonary conditions are among the leading causes of morbidity, mortality and health service use in the world \cite{wang2016global}. The diagnosis of this and other diseases targeted by chest radiographs (\textit{e.g.}, Cardiomegaly, Pneumothorax) pose a significant burden for physicians and radiologists, as this analysis can be time consuming, prone to errors and inter-expert disagreements. The computer vision-enabled interpretation of large numbers of chest X-rays has the potential to reduce the operational costs involved in the interpretation of these potentially live-saving tests. Moreover, semi-automated methodologies for pre-screening radiographs (\textit{i.e.}, flagging a sample as potentially unhealthy pending further human deliberation) expands the possible use cases of these automated analysis systems. 

A number of recent works indicate the growing interest in performing the automatic diagnosis of radiographs using computer vision \cite{dunnmon2019,Hashmi2020,Wang_2017,rajpurkar2017chexnet, stephen2019efficient, jaiswal2019identifying}. These methods are primarily focused in optimizing existing general computer vision architectures or performing transfer learning using pre-trained parameters (\textit{i.e.}, modifying only a subset of them to conform to a specific application) obtained with large image datasets. These methods often ignore optical properties of the data being used, such as its high resolution, textural features and other high-frequency components. This phenomenon is due to the common down-sampling step occurring at the input of popular CNN-based image classifiers \cite{resnet,sandler2019mobilenetv2,incept,alex,mobilenetV3}. While typically overlooked, this spatial dimension reduction may often eliminate important components of the image, in particular those of higher frequencies. Given that most down-sampling techniques use smoothing filters (\textit{e.g.}, Gaussian) to avoid aliasing, high-frequency components are lost in the increasing levels of blur. Considering that some methods reduce the dimensions of radiograph images by a factor of more than 10 \cite{dunnmon2019,Hashmi2020,Wang_2017} relative to the raw images from the CheXpert dataset \cite{irvin2019chexpert}, this issue can become particularly pressing in medical imaging.

We propose a method for preserving these finer details (high-frequency components), typically lost during the down-sampling performed before image classification. We employ a Discrete Wavelet Transform (DWT) to preserve all structural information from input images while at the same time performing the necessary down-sampling operations for the use of CNN-based classification pipelines. This can be accomplished by taking advantage of the energy-preserving properties of DWTs, as illustrated by their use in image compression schemes such as JPEG2000 \cite{JPEGstandard}. The output of this method are 3 additional independent channels to be included in both training/testing and inference phases. 
This expansion of input channels causes only a negligible increase in inference time and in the amount of memory (trainable parameters, and gradients) required for training. Increasing the number of input channels only linearly scales both FLOPs and required gradients of the network's second layer. The additional high-frequency information contained in the extra channels allow for better generalization of the dataset and ultimately increased performance.\par 

Our results show that the frequency-specific coefficients offered by the proposed method allow state-of-the-art CNNs to detect visual structures associated to frequency bands only accessible in higher-resolution images, which would otherwise be eliminated. This novel capability increases image classification performance without a relevant growth of computational load. The finer details preserved by the system proposed are particularly beneficial for medical imaging, where such high-frequency components carry important information. The remainder of this article is structured as follows. Section \ref{sec:previous_works} discusses works related to the proposed system. Section~III details our proposed approach and necessary background. In Section \ref{sec:results} we describe the experimental setup created to evaluate the proposed approach, as well as the results of experiments. Finally in Section \ref{sec:conclusion} we draw conclusions, and propose future work.    

\section{Related Works}
\label{sec:previous_works}

Related works to the proposed system include frameworks for medical image processing and classification, and energy compacting methods such as the Discrete Cosine Transform (DCT)\cite{ahmed1974discrete} and Discrete Wavelet Transforms (DWT)\cite{DWT_le1988sub} in the context of preprocessing for deep learning-based image classification.\par

\subsubsection{Deep learning-based medical image classification}
Dunnmon \etal \cite{dunnmon2019} considered 216,431 expertly annotated frontal chest radiographs and explored the ability of CNNs to correctly classify them as normal or not. The authors found using a test set of 533 samples that a relatively small number of images were required for the training of efficient general abnormality detectors: the use of 2,000 training images yielded an average area under the receiver operating characteristic curve (AUC) of 0.84, while a 20,000-sample image set generated an AUC of 0.95. This study offered a general idea on the scale expected on benchmark datasets used for medical imaging classification (in particular for radiographs). More recently Hashmi \etal \cite{Hashmi2020} proposed a system that employs transfer learning to train specialized image classifiers for the detection of Pneumonia in radiographs. Each classifier is obtained by fine-tuning pre-computed parameters from diverse CNNs (\textit{i.e.}, ResNet18\cite{resnet}, Xception\cite{chollet2017xception}, InceptionV3\cite{incept}, DenseNet121\cite{huang2018densely}, and MobileNetV3\cite{mobilenetV3}) on a dataset of chest X-ray images. The individual predictions from this ensemble of classifiers are weighted to produce the final output; this approach outperformed the predictions of any individual classifier.\par

Tang \etal \cite{Tang2020} trained multiple CNNs that scored high AUC in the distinction between normal and abnormal frontal chest radiographs (0.98 for test samples detached from the training dataset, 0.94 for samples from an external dataset). Moreover, the authors report that fine-tuning the aforementioned image classifiers using radiographs from pediatric patients resulted in an AUC of 0.944, attesting to the generalization power of the networks used. The results from Tang \etal~represent strong arguments about the fact that modern CNNs (\cite{resnet,sandler2019mobilenetv2,incept,alex,mobilenetV3}) are able to match and often surpass expert-level analysis of radiographs. 

\label{sec:methods}

\begin{figure*}[ht]
\begin{center}
\includegraphics[width=0.9\linewidth]{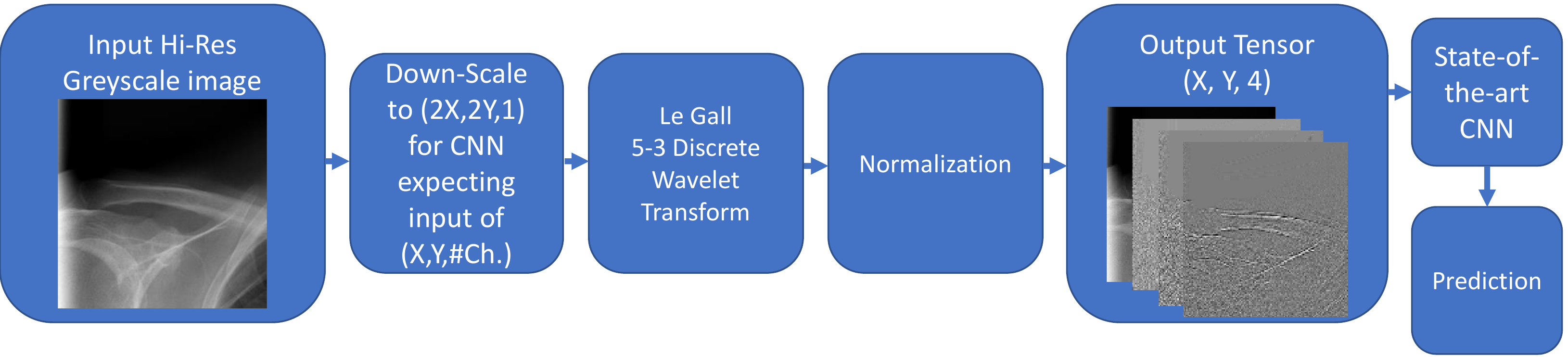}
\end{center}
   \caption{Pipeline of proposed DWT-based preprocessing method, where $(X//2,Y//2)$ yields the required input spatial dimensions for a given CNN.}
\label{fig:1-overall}
\end{figure*}

\subsubsection{DCT-Based Image energy compression preprocessing} 
Previous methods in the medical imaging field have used the frequency decomposition and signal energy compaction properties of the DCT to reduce problem dimensionality\cite{WAVLET_BRAIN,Boukhechba2018}. These methods used the DCT to refactor their input into a series of frequency bins representing channels with greatly reduced dimensions\cite{WAVLET_BRAIN,Boukhechba2018}. 
Sridlhar and Murali Krishna \cite{WAVLET_BRAIN} proposed a probabilistic neural network to detect brain tumors in magnetic resonance brain images. The images used in this work were often subject to severe noise from issues related to inconsistencies in data collection, which ultimately hindered the neural network-based classification performance. In an attempt to attenuate these issues, the authors used the Discrete Cosine Transform (DCT) to reduce image dimensionality and perform feature extraction in the frequency domain. The authors then used the DCT-based frequency features as inputs of their neural networks, instead of the original MRI images. The reduced high-frequency input noise for their approach significantly increased the model accuracy. Boukhechba, Wu, and Bazine \cite{Boukhechba2018} proposed a preprocessing approach to hyper-spectral imagery data for independent component analysis. Their method utilizes the DCT's ability to compress the energy of an input signal into a small number of frequency components. This property allowed the authors to reduce dimensionality and decrease sensitivity to high-frequency noise by only considering the lower frequency components while maintaining the majority of signal energy.

\subsubsection{DWT Image energy compression preprocessing}
Li \etal~\cite{li2020wavelet} proposed WaveCNets, which uses DWT-based replacements for the intermediate down-sampling layers within the networks. The authors replace pooling and strided convolutions with different wavelet transforms during the down-sampling phases that typical CNNs contain, continuing to consider only low frequency components. The WaveCNets models where evaluated by replacing the applicable layers in several \sate CNNs. The authors report a noticeable increase in model accuracy and tolerance to Gaussian noise in ImageNet \cite{Wang_2017} classification tasks. A wavelet-based preprocessing method for feature extraction was proposed by Reema and Babu \cite{reema} for brain tumor detection in MRI images. After applying a DWT on input MRI images, the authors extract hand-crafted visual features. The method proposed in \cite{reema} then utilizes a Support Vector Machine (SVM) to allow for the detection and segmentation of tumors based on the aforementioned features. This intial DWT-based preprocessing step proposed in \cite{reema} boosted the system performance and decreased the complexity of input MRI images used in the detection and segmentation tasks.

Our novel method utilizes similar frequency-based image decomposition to provide increased image signal energy (\textit{i.e.}, apparent resolution) to the input of CNNs commonly used for image classification. These additional information provide important visual features to the CNNs, allowing the networks to more efficiently classify radiographs and generalize for data arriving from diverse datasets. The data contained in the high-frequency components we provide, which would otherwise be lost in regular down-sampling schemes, is presented in a format that causes only a negligible increase to model size and inference time.

\section{Methods} 

Our proposed approach employs Discrete Wavelet Transforms as a means to preserve important structural information, in particular high-frequency components, of high-resolution input images before down-sampling them to conform to the dimensions required by popular CNNs (\textit{e.g.}, $224\times224$ and $299\times299$ pixels). These additional components, which would otherwise be lost, help the networks better generalize the automatic classification of radiographs (as discussed in the following Sections). Non-medical imaging applications can also take advantage of our proposed method, as illustrated by a consistent increase in performance observed over different CNNs on a dataset of generic images (\textit{i.e.}, ImageNet\cite{iamgenet}). Figure \ref{fig:1-overall} summarizes the proposed approach. \par

The only minor modification required by our framework upon existing CNNs is to increase the number of expected input channels by a factor of 4 (to accommodate the additional DWT-generated inputs). Therefore, it can be easily integrated with any system that uses CNNs as automatic feature extractors, such as image classifiers, generative adversarial networks, object detectors, instance segmentation pipelines, among others. 

\subsection{LeGall 5/3 Discrete Wavelet Transform}

Discrete Wavelet Transforms have become ubiquitous among image compression pipelines, in particular because of their higher compression ratios and lack of ``blocking'' artifacts generated by their predecessors, Discrete Cosine Transforms (DCT) \cite{angelopoulou2008implementation}. The use of 2-dimensional DWT in images creates a set number of frequency-based decomposition levels of an input, which can later be used to partially or fully reconstruct the original image (given a specific choice of filter bank).\par

\begin{figure*}[!ht]
\centering
\includegraphics[width=0.8\linewidth]{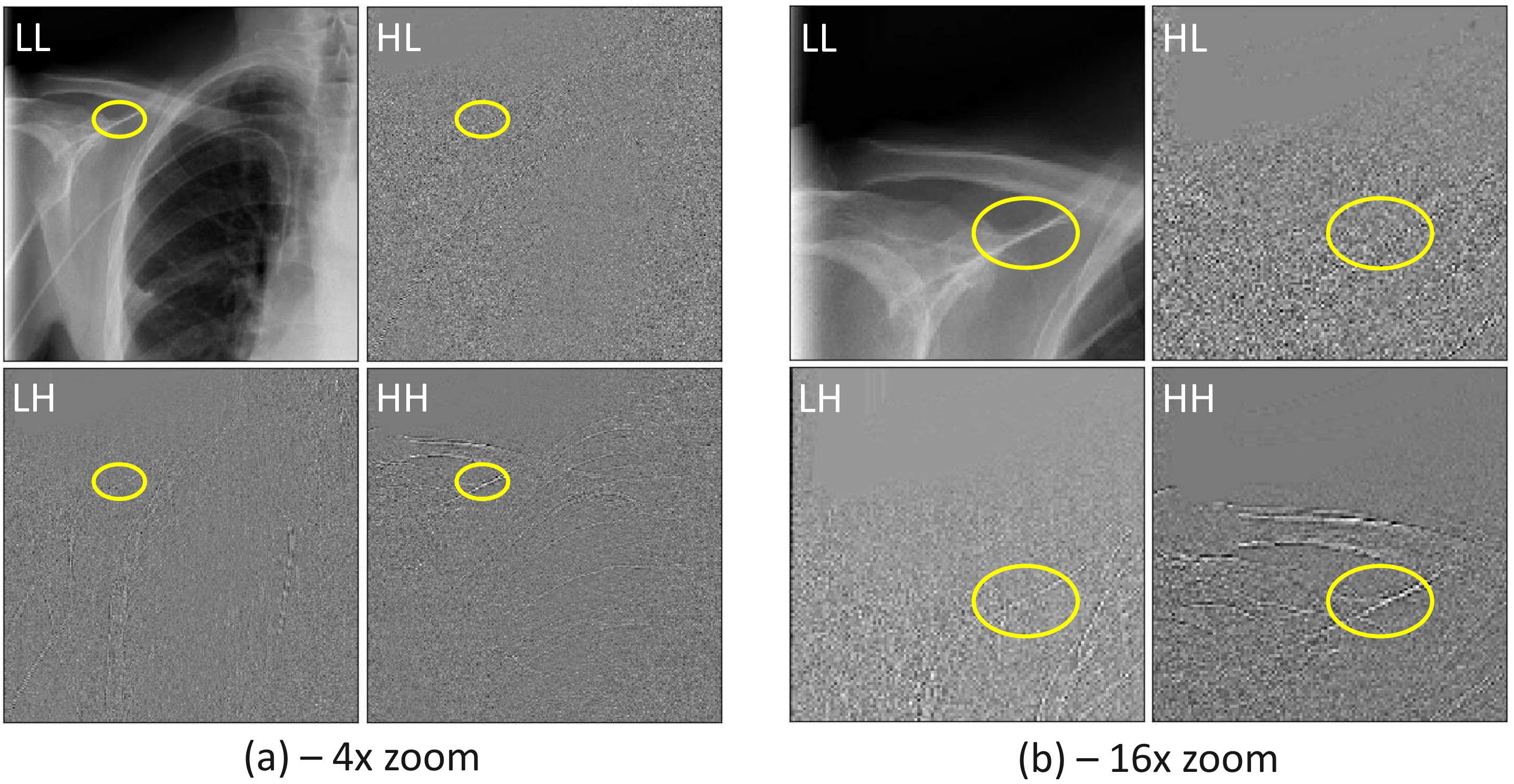}
\caption{Four frequency-based coefficients obtained with LeGall 5/3 DWT \cite{legall53} from a radiograph of the NIH Chest-8 dataset \cite{Wang_2017}. The two letters in the top left of each image indicate the filters used on the rows and columns of each input respectively. Note that the same region (yellow) of the spine of the scapula is represented in the LL (structurally equivalent to a regular down-sampled input) as well as in the HL, LH, HH. These additional channels encode all information of the input in a band-frequency manner, allowing the network to train specific visual features in each frequency represented. }\label{fig:full}
\end{figure*}

Each decomposition level created by a 2D DWT can be obtained following a two-step process (\textit{dyadic decomposition}). First, two 1D filters are convolved with only the rows of the input using a step size of two (\textit{i.e.}, centered at every other pixel index), generating two coefficients as outputs: a low- and a high-frequency one, both possessing only half the number of \textit{columns} of the input. The 1D filters proposed in LeGall 5/3 \cite{DWT_le1988sub} are detailed in Equations~(\ref{eq:f_h}) (high-frequency filter $F_h$) and (\ref{eq:f_l}) (low-frequency filter $F_l$). Second, these two coefficients are convolved with the transposes of $F_h$ and $F_l$ along their columns, further doubling the number of output coefficients to a total of four. Similarly to the first step, this convolution reduces the number of \textit{rows} by half. Note that one could choose to start the convolution operations with the columns of the input, rather than the rows.\par

\begin{equation}
\label{eq:f_h}
F_h=\{-\frac{1}{4}, \frac{1}{2}, -\frac{1}{4}\}
\end{equation}

\begin{equation}
\label{eq:f_l}
F_l=\{-\frac{1}{8}, \frac{1}{4}, \frac{3}{4}, \frac{1}{4}, -\frac{1}{8}\}
\end{equation}

\begin{table*}[b]
\scriptsize
\caption{Aggregate results of various state-of-the-art CNNs using the proposed method on the NIH Chest-8~\cite{Wang_2017} test set. Positive relative results reflect an increase in performance when using the DWT-based additional channels. Relative results of Binary Cross Entropy Loss (BCE) and Mean Squared Error (MSE) are calculated per class, then their mean is measured and reported. Number of model parameters are shown for both DWT (proposed) and grayscale-only inputs (differences are consistently less than $1\%$).}
\label{tab:chest-results}
\begin{center}
\vspace{-1.2em}
\begin{tabular}{|p{0.12  \linewidth} |p{0.145 \linewidth} | p{0.055\linewidth} | p{0.065\linewidth} |p{0.07\linewidth} |p{0.165\linewidth} |p{0.17\linewidth} |}
\hline
\textbf{CNN Model}  & \textbf{DWT (proposed method)} & \textbf{\# param.} & \textbf{Mean BCE} & \textbf{Mean MSE} & \textbf{Mean Relative Change BCE} &  \textbf{Mean Relative Change MSE} \\
\hline

MobileNetV2 \cite{sandler2019mobilenetv2} & N & 2.24M & 0.648 & 0.228 & &\\ 
MobileNetV2 \cite{sandler2019mobilenetv2} & Y & 2.24M & 0.680 & 0.239 & -4.85\% & -5.10\%\\ \hline
AlexNet \cite{alex} & N &  25.4M & 0.721 & 0.262 &  & \\ 
AlexNet \cite{alex} & Y &  25.4M & 0.712 & 0.258 & {1.45}\% & {1.79}\%\\ \hline
ResNet50 \cite{resnet} & N &  23.6M & 0.621 & 0.217 &  & \\ 
ResNet50 \cite{resnet} & Y &  23.6M & 0.569 & 0.192 & {6.49}\% & {8.42}\%\\ \hline
InceptionV3 \cite{incept} & N &  21.8M & 0.583 & 0.199 &  & \\
InceptionV3 \cite{incept} & Y &  21.8M & 0.556 & 0.187 & 1.32\% & {1.83}\%\\
\hline
SquezeNetV1.1 \cite{iandola2016squeezenet} & N &  1.24M & 0.634 & 0.231 &  & \\
SquezeNetV1.1 \cite{iandola2016squeezenet} & Y &  1.24M & 0.611 & 0.219 & 2.98\% & {3.86}\%\\

\hline
\end{tabular}
\end{center}
\vspace{-2em}
\end{table*}

The input image is typically referred to as $LL_0$, and the outputs of the first application of $F_h$ and $F_l$ as $H_0$ and $L_0$, respectively. While $L_0$ represents an approximation of the input signal at a coarser resolution, $H_0$ carries the high-frequency details of it \cite{angelopoulou2008implementation}. The second application of $F_h$ and $F_l$, this time in the columns of $L_0$ and $H_0$, creates four outputs, namely: $LL_1$, $HL_1$, $LH_1$ and $HH_1$ (the subscript reflects the decomposition level). These four outputs possess half the spatial resolution of the input $LL_0$.\par 

We seek to preserve the high-frequency components of the input images during down-sampling, and a one-level \textit{dyadic}, DWT-based decomposition generates four coefficients with half the spatial resolution of the input. Therefore, in order to match the size of such coefficients with the required input dimensions from CNNs, we first resize the input images using bi-linear interpolation to double the resolution required by a network (\textit{e.g.}, $598\times598$ for a requirement of $299\times299$), and continue to apply the DWT-based decomposition. As a result we match the input size requirements of CNNs while preserving all information from the twice-spatial resolution input. Figure \ref{fig:1-overall} illustrates this process. \par

The frequency-based coefficients $LL_1$, $HL_1$, $LH_1$ and $HH_1$ are used as the input channels for the image classification CNNs that we train. Figure \ref{fig:full} illustrates the four decomposition components discussed (note that we suppress the subscript because only one level is used).

The LeGall 5/3 filters create a lossless representation of their inputs \cite{DWT_le1988sub}, meaning that an input image can be recovered using exclusively the calculated  decomposition coefficients. This lossless reconstruction capability is especially useful in our work because all information is carried in the inputs forming a 4-channel representation, each with half of the spatial resolution of the original images. This group of channels creates a frequency-based separation between the components of the input, encoding structurally independent but semantically related features. Based on the results presented in the remainder of this article, we hypothesize that providing frequency-specific coefficients (\textit{e.g.}, low-frequency or high-frequency only; see Figures \ref{fig:full} (a) and (b)),  allows the CNNs to specialize in the identification of structures specific of a frequency band; this ultimately increases classification performance, without an increase in computational load.\par

Figure~\ref{fig:full} shows that the typical down-sampled input (``LL'' images of this Figure) is enhanced with three additional channels that better characterize high-frequency regions, such as the clavicle edges highlighted in yellow. These complementary inputs encode all information (that would otherwise be lost in down-sampling), allowing for a better generalization of the CNNs that use these coefficients.  

\subsection{Considerations on Wavelets and Image Channel Encoding}

The frequent scaling operations of modern CNNs represent a critical factor in determining the allocation of resources: any increase in spatial resolution by a factor of $X$ increases the memory required for training and the FLOPs of the convolutional layers by the same factor $X$. Due to this scaling and limited memory and computation resources, the input image resolution is often severely limited.  \par

Our proposed approach uses LeGall 5/3-based complementary inputs (\textit{i.e.}, coefficients of the dyadic decomposition) in the form of three extra input channels to encode all the information from the original image (despite the reduction in spatial resolution), as illustrated in Figure~\ref{fig:1-overall}. This strategy, different from simply providing higher-resolution images and dealing with their resource implications, only negligibly increases the number of FLOPs and memory required. In fact, only the number of parameters of the first layer in a CNN is increased. For instance, the three extra DWT-based channels we propose only increased the number of parameters in a MobileNet \cite{mobilenetV3} by $0.02\%$ (see other examples on Tables~\ref{tab:chest-results} and \ref{tab:imagenet-results}). This percentage is even smaller for bigger networks (\textit{e.g.}, ResNet50~\cite{resnet}, InceptionV3~\cite{incept}). As a result, our method provides all the information from an image with twice the spatial resolution required by CNNs, but instead of doubling the number of parameters involved in their training, it only very slightly increases it. \par

A detailed analysis of the 4-channel input we provide (see Figure~\ref{fig:full}) follows. 

\begin{enumerate}
    \item The first channel, $LL$, which represents the result of applying two low-pass filters ($F_l$ and its transpose) to rows and columns of the input, is structurally similar to the a regularly down-sampled image (often obtained using nearest neighbor, bi-linear or bi-cubic interpolation).
    \item The following three channels, $HL$, $LH$ and $HH$ are a combination of the outputs of $F_l$ and $F_h$, or singularly $F_h$ and its transpose. As detailed in \cite{legall53}, the coefficients obtained with the application of filters $F_l$ and $F_h$ can lead to a perfect reconstruction of the input by representing only specific frequency bands at a time (\textit{e.g.}, ``$HL$: only the lowest frequencies from the high-frequency components of the input'').         
\end{enumerate}

Although we consider only one-channel images (\textit{i.e.}, grayscale) in this work, our method can be extended to multi-channel images. For example, a 3-channel RGB image could be represented by 12 DWT-generated coefficients. \par

\section{Experimental results} 
\label{sec:results}

\subsection{Experimental Setup}
We utilize several state-of-the-art CNNs to evaluate our proposed method with respect to compatibility and scaling of different models' parameters. We use the same hyper-parameters (specified below) in the training phases of all models for consistency. For each CNN we train the image classifier with wavelet-supplemented inputs (\textit{i.e.}, the proposed approach) and regular ones. \par
In order to evaluate the potential of the proposed approach in medical and natural images alike, we perform our experiments in two large datasets: NIH Chest-8 Dataset~\cite{Wang_2017}, composed of $108,000$ de-identified and annotated images of chest X-rays, and ImageNet-2017~\cite{iamgenet}, which possesses $1,200,000$ images spanning 1,000 categories. ImageNet-2017 was chosen as it allows for an analysis of the generalization capabilities of the proposed method in a large dataset of diverse natural images. \par
The models involved in our analysis are MobileNetV2 \cite{sandler2019mobilenetv2}, ResNet-50 \cite{resnet}, AlexNet \cite{alex}, and InceptionNet \cite{incept} on the NIH Chest-8 Dataset \cite{Wang_2017}, and SqueezeNet \cite{iandola2016squeezenet}, MobileNetV2 \cite{sandler2019mobilenetv2}, ResNet-50 \cite{resnet}, and InceptionNet \cite{incept} on our ImageNet-2017 benchmark \cite{iamgenet} of grayscale-only images. We do not train AlexNet \cite{alex} on ImageNet because of its large number of parameters and because it has been shown that more recent CNNs (\textit{e.g.}, \cite{resnet,incept}) outperform it.  


\subsection{Considerations on the ImageNet-2017 \cite{iamgenet} dataset}
While training/testing we performed classification on all 1,000 classes of the dataset. Images are converted to grayscale for training and evaluation (as discussed in the previous Section). The hyper parameters used for all models and ImageNet were: categorical cross entropy loss, spatial input dimensions of $224\times224$ pixels, learning rate of $1e^{-4}$, batch size of 32, ADAM optimizer \cite{kingma2014adam} and an indefinite number of epochs (the training stopped when the validation loss for each model stabilized). 

\subsection{Considerations on the NIH Chest-8 Dataset~\cite{Wang_2017} dataset} 
We modify the multi-class classification task proposed by the dataset to eight separate binary classification tasks, one for each class. This allows for more points of comparison between wavelet- and non-wavelet-based inputs. The eight classes of the NIH Chest-8 dataset are: Pneumothorax, Effusion, Mass, Pneumonia, Cardiomegaly, Nodule, Atelectasis, and Infiltration. To handle the data imbalance for our new problem formulation we randomly sample a number of negative samples from the test set based on the number of positive samples creating a ratio of $1:1$ per class-specific classification task. \par

Testing and validation on the NIH Chest-8 Dataset were performed on the entire testing and validation subsets, respectively. Weights are initialized using the randomized method proposed by He~\etal~\cite{he2015delving}. The hyper parameters used for all models trained on this dataset were: binary cross entropy loss (BCE), spatial input dimensions of $512\times512$ pixels, learning rate of $1e^{-5}$, batch size of 8, ADAM optimizer \cite{kingma2014adam} and an indefinite number of epochs (again, training stopped when the validation loss for each model stabilized). 

\begin{table*}[!tp]
\scriptsize
\caption{Results from various state-of-the-art CNNs on grayscale ImageNet-2017~\cite{iamgenet} validation set. Positive relative results reflect an increase in performance when using the DWT-based additional channels (proposed). Number of model parameters are shown for both DWT (proposed) and grayscale-only inputs (differences are consistently less than $1\%$)}
\label{tab:imagenet-results}
\begin{center}
\vspace{-1.5em}
\begin{tabular}{| p{0.12\linewidth} | p{0.10\linewidth} | p{0.08\linewidth} |p{0.06\linewidth} |p{0.06\linewidth} |p{0.06\linewidth} | p{0.1\linewidth} | p{0.1\linewidth} | p{0.1\linewidth} |}
\hline
\textbf{Model}  & \textbf{DWT  (proposed method)} & \textbf{\# params} & \textbf{CCE} & \textbf{Top-1} & \textbf{Top-5} & \textbf{Relative Change CCE} & \textbf{Relative Change Top-1} & \textbf{Relative Change Top-5}\\
\hline

MobileNetV2 \cite{sandler2019mobilenetv2} & N & 3.56M &  2.08 & 53.0\%  & 77.6\% &  & &  \\

MobileNetV2 \cite{sandler2019mobilenetv2} & Y & 3.56M & 2.051 & 53.4\% & 78.0\% & 1.49\% & 0.75\%  & 0.52\%\\\hline

SquezeNetV1.1 \cite{iandola2016squeezenet} & N & 1.24M & 2.81 & 43.4\% & 68.8\% &  &  & \\

SquezeNetV1.1 \cite{iandola2016squeezenet} & Y & 1.24M & 2.72 & 45.3\% & 69.9\% & 3.16\% & 4.37\% & 1.60\% \\\hline

ResNet50 \cite{resnet} & N & 25.6M & 2.23 & 51.2\% & 76.1\% &  &  &   \\

ResNet50 \cite{resnet} & Y & 25.6M & 2.16 & 52.5\% & 77.3\% & 3.23\% & 2.54\% & 1.58\%\\\hline

InceptionV3 \cite{incept} & N & 23.9M & 1.97 & 55.4\%  & 79.5\% &  &   & \\

InceptionV3 \cite{incept} & Y & 23.9M & 1.93 & 57.1\% & 80.9\% & 1.83\% & 3.16\% & 1.76\%\\\hline

\hline
\end{tabular}
\end{center}
\vspace{-2em}
\end{table*}

\subsection{NIH Chest-8 Results}

Due to the reformulation of the NIH Chest-8 task to a series of 8 binary classification problems, the ratio of positive to negative examples in the test set exceed $1:50$ for some classes. For this reason, the utilization of metrics based on True and False Positives would not result in useful performance insights. Instead we report BCE and mean squared error (MSE) values for the analysis of the predictions of each model. \par

Table~\ref{tab:chest-results} provides the aggregate relative performance gains or losses between the wavelet pre-processed inputs (proposed method) and regular ones considering the five tested CNNs. These results reflect an aggregate measure over the test set for all 8 binary classification problems (\textit{i.e.}, a binary classification per class of the NIH Chest-8 Dataset). Table~\ref{tab:chest-results} shows that the proposed method resulted in improvements (\textit{i.e.}, reduction) in the test set BCE loss for all models but MobileNetV2.\par

We hypothesize that the MobileNet results are related to two main reasons, namely:
\begin{enumerate}
    \item MobileNet's early spatial down-sampling (particular to this CNN architecture) nullifies the additional information provided by the proposed method before features can be properly extracted/learned.    
    \item Non-ideal choice of hyper parameters for this model. While other methods could more quickly generalize the training data, this smaller model was not able to do that for the same set of hyper parameters (\textit{e.g.}, potentially due to the learning rate being too large, preventing the model to reach convergence).
\end{enumerate}

Tested models excluding MobileNetV2\cite{sandler2019mobilenetv2} (\textit{i.e.}, ResNet-50 \cite{resnet}, AlexNet \cite{alex}, InceptionNet \cite{incept}, and SquezeNetV1.1 \cite{iandola2016squeezenet}) performed better by an average of $3.1\%$ on the test set of the NIH Chest-8 dataset when considering test BCE and using the proposed method. This indicates that these models, with the same set of hyper parameters, were able to use the higher-frequency information encoded in the additional channels created in our method to improve their generalized predictions.\par

This improvement in performance is most likely due to the additional information encoded in the three extra channels our method provides, which would otherwise be lost in regular down-sampling pipelines (\textit{e.g.}, high-frequency portions of a specific texture that would be eliminated with a simple resizing of the image).\par

Ablation studies summarized in Table~\ref{tab:chest-ablation-results} show that only when using all three DWT-based additional inputs the performance of the different models on the NIH Chest-8 increased. This indicates that the additional high-frequency components of our method do not just present new individual interest points (\textit{e.g.}, edges and corners) to the network, as the output of a Sobel or Laplacian of Gaussian (LoG) operator would, but rather the entirety of the information present in the input. This signal-energy-preserving characteristic appears to be necessary for efficient representation of coherent statistical features that can be learned by the CNNs. 

\begin{table}[b]
\scriptsize
\caption{Ablation study results for ResNet50~\cite{resnet} on NIH Chest-8 Dataset\cite{Wang_2017} showing the influence of using different combinations of the DWT-based channels.}
\label{tab:chest-ablation-results}
\begin{center}
\vspace{-1.5em}
\begin{tabular}{|p{0.05  \linewidth} |p{0.05  \linewidth} | p{0.05\linewidth} | p{0.05\linewidth} |p{0.15\linewidth} |p{0.15\linewidth} |}
\hline
\textbf{LL} &\textbf{HL} &\textbf{LH} &\textbf{HH} &\textbf{Mean BCE} & \textbf{Mean MSE} \\
\hline

Y & N & Y & Y & 0.942 & 0.352\\ \hline
Y & Y & N & Y & 0.969 & 0.362\\ \hline
Y & Y & Y & N & 0.963 & 0.359\\ \hline
Y & N & N & N & 0.621 & 0.217\\ \hline
Y & Y & Y & Y & \textbf{0.569} & \textbf{0.192}\\ \hline

\hline
\end{tabular}
\end{center}
\end{table}

\subsection{ImageNet-2017 Results}
The heterogeneity of this dataset distinguishes it from the more semantically coherent images found in the previously discussed NIH Chest-8 dataset. Thus it represents an efficient indicator of the usefulness of the proposed method for heterogeneous visual data.\par

Table~\ref{tab:imagenet-results} shows that we obtained results in the ImageNet-2017 experiments which are similar to those using the NIH Chest-8 dataset, with an average relative increase of $2.42\%$ in performance across all models for Categorical Cross Entropy (CCE) and $2.71\%$ relative increase in Top-1 Accuracy. Notably, in this evaluation all models show an increase in relative performance due to the additional high-frequency features encoded by the proposed method. This likely happens because the classes of ImageNet are more easily differentiated with the use of high-frequency information, for instance the highly divergent textures between the \textit{dog} and \textit{boat} classes.  

Note that the results from ImageNet-2017 are under-reporting the contributions of proposed method, given that the majority of images provided by the dataset fall below the minimum ideal source resolution of $448\times448$ pixels (double the input dimensions of a given CNN). Regardless, results indicate a systematic increase in classification performance across all state-of-the-art CNNs evaluated stemming from the use of the proposed DWT method. \par 

We expect to observe further improvements in classification performance for datasets that contain higher-resolution images. The results on this large dataset show that the additional high-frequency coefficients offered by the proposed method contribute to the classification of images of diverse natures (as illustrated by the $1,000$ visual classes of ImageNet-2017\cite{iamgenet}).

\begin{figure}[!b]
    \centering
    \includegraphics[width=0.75\linewidth]{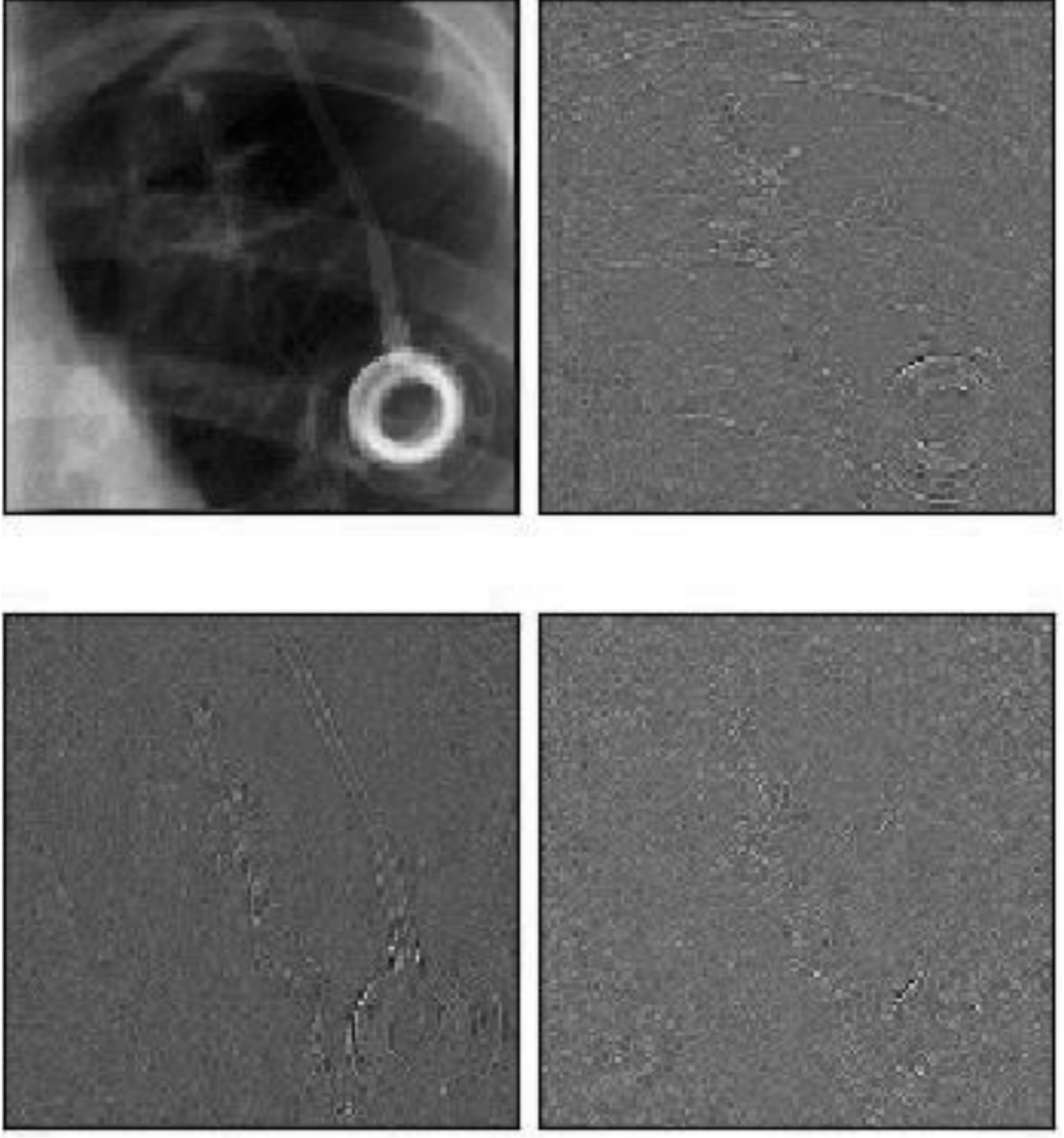}
    \caption{Illustrative sample from the Mass class (NIH Chest-8~\cite{Wang_2017} dataset) and its DWT-based inputs. Cropped based on detection annotation.}\label{fig:mass}
\end{figure}

\subsection{Class-specific Influence of DWT-based Inputs}

In the previous sub-sections the influence of the additional high-frequency DWT-based channels were presented on aggregate metrics considering all classes. Here we focus on the relative class-specific improvements. The worst-performing class with the use of our method in the NIH Chest-8 dataset (when compared to down sampled inputs) was Cardiomegaly.\par

\begin{figure}[t]
  \centering
  \includegraphics[width=0.8\linewidth]{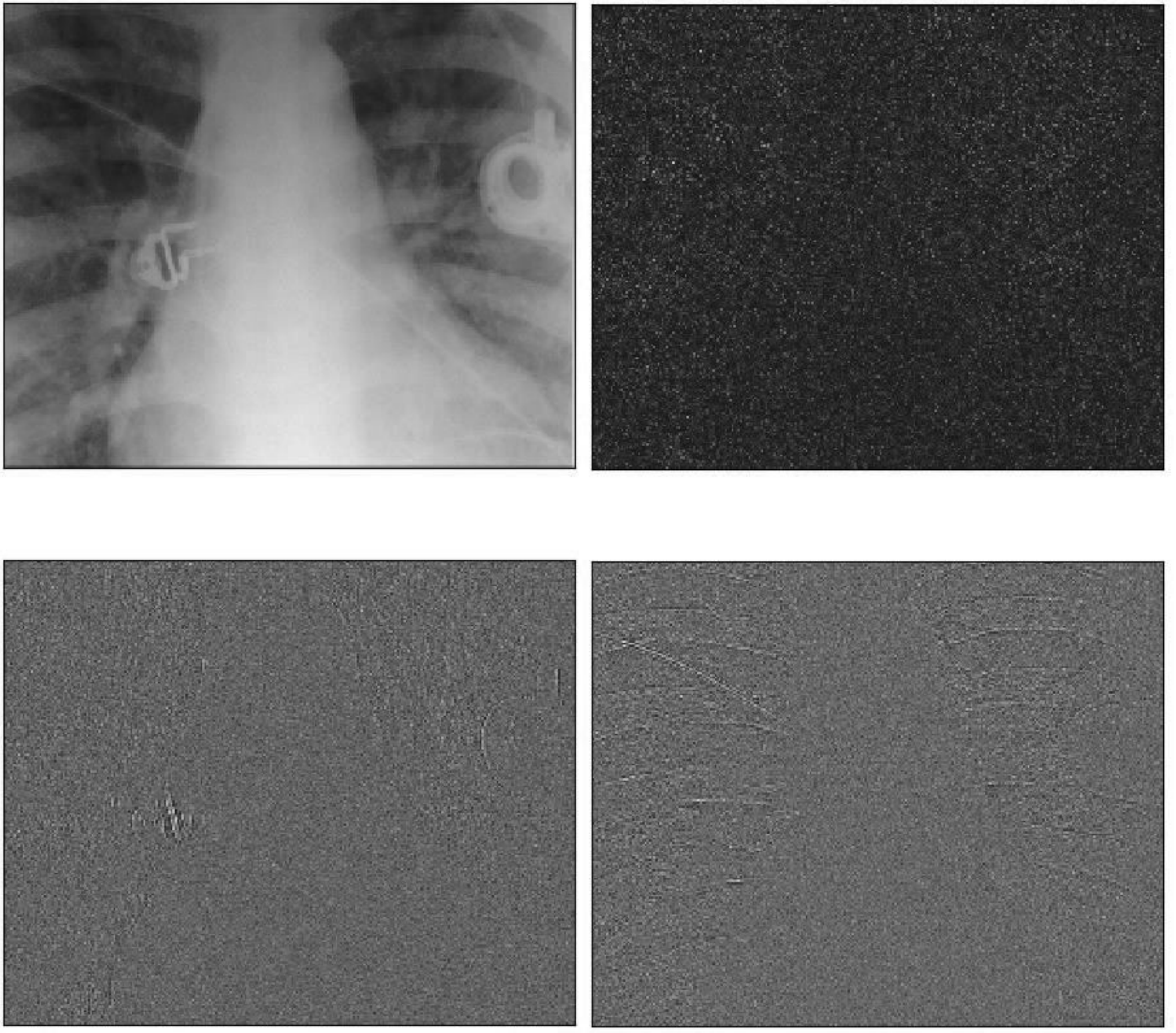}
  \caption{Illustrative sample from the Cardiomegaly class (NIH Chest-8~\cite{Wang_2017} dataset) and its DWT-based inputs. Cropped based on detection annotation.}\label{fig:cardio}
  \vspace{-1.5em}
\end{figure}

Cardiomegaly is characterized in radiographs as an enlarged heart (see Figure~\ref{fig:cardio}). The relevant visual features used in the classification of Cardiomegaly are concentrated in lower frequencies of the image (\textit{e.g.}, the total size of homogeneous regions of the heart is an important visual cue for this class).  
As discussed, our DWT-based method preserves additional information in higher frequencies. Therefore, not only these additional high-frequency components are not necessary for an efficient classification of Cardiomegaly, they might represent a detrimental addition (\textit{i.e.}, high-frequency noise). 

Previously in the NIH Chest-8 dataset benchmarks Mass was noted to be a particularly challenging class \cite{Wang_2017}, due to its intraclass variability. Our experiments reported an increase in classification performance in this class when using the proposed DWT-based inputs. Such increase can be associated to the fact that edges and changes in textures (high-frequency components) represent the main visual cues of this class (see Figure~\ref{fig:mass}), therefore its classification is particularly benefited by the additional inputs we provide. This phenomenon is illustrated by the high-frequency structural elements that are easily visualized in the DWT-based channels of Figure~\ref{fig:mass}, in contrast with those from Figure~\ref{fig:cardio}.\par 

Based on our thorough experiments and ablation studies we conclude that the additional inputs provided by the proposed method are effective in offering useful supplementary high-frequency components for classification of images, in particular for classes where fine features and textures are particularly distinctive.

\section{Conclusion} 
\label{sec:conclusion}
We propose a novel and simple, CNN-agnostic method that encodes all information from higher-resolution images when down-sampling them for use with networks of image classification. These extra information are represented as three additional channels (per original channel of the input), and our experiments show that they systematically increase the performance of state-of-the-art CNNs for image classification.\par 

Our proposed method performs a Discrete Wavelet Transform (DWT) using the LeGall 5/3 filter bank~\cite{legall53} to encode additional high-frequency components of the input, effectively presenting all of its information after down-sampling. These complementary channels add only a negligible number of parameters to the networks and require minor modifications to the first few layers of a CNN architecture, thus representing an approach that can be easily incorporated to existing systems. These supplemental channels allow diverse CNNs to better detect visual structures that are associated with specific frequency bands otherwise lost. This improvement is highlighted in the results we present: our method improves the image classification performance of various CNNs without a noticeable increase in computational load. \par

In tested models excluding MobileNetV2\cite{sandler2019mobilenetv2} our proposed approach reduced testing loss (BCE) by $3.1\%$ in binary classification tasks from the NIH Chest-8 dataset~\cite{Wang_2017}. Detailed analysis of the interaction between the proposed method and the \textit{Cardiomegaly} and \textit{Pneumothorax} classes of \cite{Wang_2017} revealed that these additional channels are particularly beneficial to classes where texture or finer details are predominant. \par

Beyond medical imagining usage, we explore the potential of the proposed approach in a large generic dataset, ImageNet-2017~\cite{iamgenet}. Considering the performance on all $1,000$ classes of this dataset, our experiments show an average relative decrease of $2.42\%$ across all CNNs models evaluated for Categorical Cross Entropy loss and $2.71\%$ relative increase in Top-1 Accuracy. These improvements highlight how the high-frequency components offered by our method can help CNNs better generalize for medical imaging and generic datasets alike. \par

Future work will address the application of DWT using other wavelets for data encoding, as well as investigate the effectiveness of these additional input channels to systems that use CNNs as features extractors for other tasks, such as: object detection, instance segmentation and Generative Adversarial Networks-based applications.

\section*{Acknowledgments}

The authors would like to acknowledge the valuable contribution received by the \textit{Jamie Cassels Undergraduate Research} award \footnote{www.uvic.ca/learningandteaching/students/student-awards/jcura/index.php}, and the University of Victoria for their logistical and academic support. 

\bibliographystyle{IEEEtran}
\bibliography{egbib}
\end{document}